\documentclass[aps,prd,twocolumn,superscriptaddress,nofootinbib,10pt]{revtex4-1}
\usepackage[paperwidth=21cm,paperheight=29.7cm,top=2.54cm,bottom=2.54cm,left=2cm,right=2cm]{geometry}
\usepackage[colorlinks=true,allcolors=blue]{hyperref}
\usepackage{graphicx,color}
\usepackage{dcolumn}
\usepackage{subfigure}
\usepackage{bm}
\usepackage{amsmath,amsfonts,amssymb}
\usepackage{acronym}
\usepackage{enumitem}
\usepackage{array}
\usepackage{multirow}
\allowdisplaybreaks
\newcommand{\be}{\begin{equation}}
\newcommand{\ee}{\end{equation}}
\newcommand{\bea}{\begin{eqnarray}}
\newcommand{\eea}{\end{eqnarray}}

\newcommand{\Msun}{{\rm M}_\odot}
\newcommand{\dd}{\mathrm{d}}
\newcommand{\Rin}{R^{\textup{in}}_{\ell m \hat{\omega}}} 
\newcommand{\Rout}{R^{\textup{out}}_{\ell m\hat{\omega}}} 

\newcommand{\SOUTHCUT}{
School of Physics and Optoelectronics, South China University of Technology, Guangzhou 510641,
People's Republic of China}

\newacro{EMRI}{Extreme Mass Ratio Inspiral}
\newacro{MBH}{massive black hole}
\newacro{BH}{black hole}
\newacro{GR}{general relativity}
\newacro{HKBH}{hairy Kerr black hole}
\newacro{KBH}{Kerr black hole}
\newacro{NHT}{no-hair theorem}
\newacro{DWD}{double white dwarf}
\newacro{GW}{gravitational wave}
\newacro{AK}{analytic kludge}
\newacro{NK}{numerical kludge}
\newacro{AAK}{augmented analytic kludge}
\newacro{CO}{compact object}
\newacro{PE}{parameter estimation}
\newacro{SNR}{signal-to-noise ratio}
\newacro{PN}{post newtonion}
\newacro{FIM}{Fisher information matrix}
\newacro{LSO}{last stable orbit}
\newacro{ISCO}{innermost stable circular orbit}
\newacro{BBH}{Binary Black Hole}
\newacro{BNS}{Binary Neutron Star}
\newacro{NS}{Neutron Star}
\newacro{KN}{Kerr-Newmann}
\begin{document}
\title{Gravitational waves from extreme mass ratio	inspirals around a hairy Kerr black hole}

\author{Tieguang Zi}
\email{zitg@scut.edu.cn}
\author{Peng-Cheng Li}
\email{pchli2021@scut.edu.cn, Corresponding author}

\affiliation{\SOUTHCUT}

\begin{abstract}
 Recently, Contreras et al. \cite{Contreras:2021yxe} introduced  a  new type of black hole, called hairy Kerr black hole (HKBH), which describes a Kerr BH surrounded by an axially symmetric fluid with conserved energy momentum tensor. In this paper, we compute the gravitational waves emitted from the extreme mass ratio inspirals around the HKBHs. We solve the Dudley-Finley equation, which describes the gravitational perturbations of the HKBH, and obtain the energy fluxes induced by a stellar-mass compact object moving on the equatorial, circular orbits. Using the adiabatic approximation, we evolved the radii of the circular orbits by taking into account the backreaction of gravitational radiation. Then we calculate the dephasing and mismatch of the EMRI waveforms from the HKBH and Kerr BH  to assess the difference between them. The results demonstrate that the EMRI waveforms from the HKBH with deviation parameter larger than $0.001$ and hair charge smaller than $1.5M$  can be discerned by LISA.
\end{abstract}

\maketitle
\section{Introduction}
 As one of the most important prediction of \ac{GR}, the existence of the \ac{BH} has been verified in the realm of
\ac{GW} observation \cite{LIGOScientific:2016aoc,LIGOScientific:2018mvr} and the direct images of supermassive BHs
\cite{EventHorizonTelescope:2019dse,EventHorizonTelescope:2019ths,EventHorizonTelescope:2022wok,EventHorizonTelescope:2022xqj}.
According to the no-hair theorem, it is believed that astrophysical BHs  are described by the Kerr-Newman (KN) family, which is uniquely characterized by the mass, spin and charge \cite{Kerr:1963ud,Newman:1965my}.
However, BHs enable to carry hair owing to the present of global charges within the frame of \ac{GR} (see a review \cite{Herdeiro:2015waa}). In general, hairy BHs  are supposed to be stationary solutions with nontrivial fields, which include scalar hair \cite{Sotiriou:2013qea,Herdeiro:2014goa,Kleihaus:2015aje}
or proca hair \cite{Herdeiro:2016tmi}.
J. Ovalle developed the
gravitational decoupling (GD) approach \cite{Ovalle:2017fgl,Ovalle:2018gic} to describe the deformations of known spherically symmetric solutions in GR induced by additional sources such as dark matter.
Recently, this method  was extended to the axially symmetric case and a rotating non-Kerr BH called hairy Kerr black hole (HKBH) was obtained \cite{Contreras:2021yxe}. Since the introduction of this solution, there have emerged several works to study the strong-field physics near the HKBH, including gravitational lensing \cite{Islam:2021dyk},
BH shadow \cite{Afrin:2021imp},  Lense-Thirring effect \cite{Wu:2023wld} and quasinormal modes \cite{Cavalcanti:2022cga,Li:2022hkq,Yang:2022ifo,Avalos:2023jeh}.

The increasing observations of \ac{GW} events from binaries since GW150914 \cite{LIGOScientific:2016aoc} have
launched a revolution of studying strong  gravitational field near the compact objects.
Till now,  ones fail to find the evidence supporting the new physics beyond \ac{GR} according to these current testing conclusions with observation of GW from stellar-mass binaries \cite{LIGOScientific:2019fpa,LIGOScientific:2018dkp,LIGOScientific:2021sio,EventHorizonTelescope:2022xqj}.
The remarkable characteristic of a BH is the existence of horizon, however, there is lack of the direct observational smoking gun in support of horizon \cite{Cardoso:2019rvt}. On the other hand, it may be difficult to rule out non-Kerr BHs
with the current observational results
\cite{Will:2005va,LIGOScientific:2019fpa,LIGOScientific:2021sio,EventHorizonTelescope:2020qrl,EventHorizonTelescope:2022xqj,Khodadi:2021gbc}.
The next generation ground-based  and space-borne GW detectors, such as Einstein Telescope \cite{Sathyaprakash:2012jk},
LISA \cite{LISA:2017pwj}, TianQin \cite{TianQin:2015yph,Gong:2021gvw} and Taiji \cite{Hu:2017mde},
are predicted to observe more kinds of signals with higher \ac{SNR} emitted from more massive objects.
Ones would have the opportunity to test the gravitational theories near the horizon with the unprecedented precision,
then it is expected that search for the smoking gun of new fundamental physics with the projected GW detectors \cite{Berti:2015itd,LISA:2022kgy,LISA:2022yao}.

Binary objects with the smaller mass-ratio lurking in the center of galaxies is evaluated as one of key target sources of the space-borne GW detectors, whose members contain a \ac{MBH} with $M\sim [10^5,10^7]\Msun$ and   stellar-mass \ac{CO} with $\mu\sim [1,10^2]\Msun$. Such an extreme mass ration inspiral (EMRI) system will emit GW signals with frequencies $\sim0.01\rm Hz$. The \ac{CO} generally  accumulates $\sim M/\mu$ orbital cycles around the \ac{MBH}, thus the EMRI event is expected to be an ideal laboratory of testing the \ac{GR} and alternatives in the region of strong field \cite{Berry:2019wgg,LISA:2022kgy,Babak:2017tow,Zi:2021pdp,Zi:2022hcc,Shen:2023pje}.
Since the EMRI dynamic is involved in the strong field and long inspiral time, there have appeared EMRI waveform models with various  physical effects or modified gravitational theories using different method, such as the Kludge models to compute waveform rapidly \cite{Barack:2003fp,Babak:2006uv,Chua:2017ujo}, the full relativistic EMRI waveform model including order-reduction and deep-learning techniques \cite{Chua:2020stf,Katz:2021yft} and adiabatic waveform model using numerical approach \cite{Hughes:2021exa} and analytical approach \cite{Isoyama:2021jjd}.
In addition there were some EMRI waveform models in the alternatives of gravity \cite{Barausse:2006vt,Sopuerta:2009iy,Pani:2011xj,Gair:2011ym,Canizares:2012is,Yunes:2011aa,Zi:2023pvl},
fundamental fields \cite{Maselli:2020zgv,Maselli:2021men,Barsanti:2022vvl,Liang:2022gdk} and the waveform models modified by  astrophysical environments \cite{Cardoso:2022whc,Figueiredo:2023gas,Speri:2022upm,Dai:2023cft}.

In this paper, we intend to compute the waveforms from the EMRI around the \ac{HKBH} constructed by the gravitational decoupling method. The EMRI dynamics is simulated with the perturbation theory of BH due to the axially symmetry.
We calculate the energy fluxes by solving the modified Teukolsky equation in term of the Dudley-Finley equation \cite{Kokkotas:1993ef,Berti:2005eb}, then evolve the orbital parameter under radiation reaction to obtain the trajectories.
To assess the phase difference between EMRI waveforms from the \ac{HKBH}  and the \ac{KBH}, we will calculate dephasing and mismatch.


The paper is organized as follows. In Sec. \ref{Method}, we introduce the method of computing EMRI waveform,
the following subsections introduced briefly some specific procedures,
subsection  \ref{background} is the spacetime background of hairy Kerr black hole,
the perturbation equations is introduced in subsection \ref{perturbeqs},  the method of adiabatic evolution is introduced in subsection \ref{evolution} and  the method of waveform data analysis is introduced in subsection \ref{waveformanalysis}.
Finally, we present the conclusion in section \ref{Conclusion}. Throughout this paper, we use the geometric units $G=c=1$.
\section{Method}\label{Method}
\subsection{Spacetime background}\label{background}
In this subsection, we firstly present a terse review of the construction of the HKBH via the GD approach \cite{Contreras:2021yxe} and then discuss the basic properties of the solution. 

Start with the Einstein equation 
\be\label{Einsteinequation}
\tilde{G}_{\mu\nu}=\tilde{R}_{\mu\nu}-\frac{1}{2}\tilde{R} \tilde{g}_{\mu\nu}=8\pi \tilde{T}_{\mu\nu},
\ee
with the energy-momentum tensor containing two
contributions,
\be\label{energymomentumtenosr}
\tilde{T}_{\mu\nu}=T_{\mu\nu}+S_{\mu\nu},
\ee
where $T_{\mu\nu}$ is associated with a known solution of
general relativity, whereas $S_{\mu\nu}$ may contain new fields or a
new gravitational sector. 

The authors of \cite{Contreras:2021yxe} considered a simple extension of the Kerr metric, which happens to be a rotational version of a Kerr-Schild spherically symmetric space-time, that is 
\bea\label{metricansatz}
ds^2&=&-\left[1-\frac{2r \tilde{m}(r)}{\tilde{\rho}}\right]dt^2-\frac{4\tilde{a}r\tilde{m}(r) \sin ^2\theta}{\tilde{\rho}^2}dtd\phi \\
&&+\frac{\tilde{\rho}^2}{\tilde{\Delta}}dr^2+\tilde{\rho}^2d\theta^2+\frac{\tilde{\Sigma} \sin^2\theta}{\tilde{\rho}^2}\phi^2,
\eea
where 
\be
\tilde{\rho}^2=r^2+\tilde{a}^2\cos^2\theta,
\ee
\be
\tilde{\Delta}=r^2-2r\tilde{m}(r)+\tilde{a}^2,
\ee
\be
\tilde{\Sigma}=(r^2+\tilde{a}^2)^2-\tilde{a}^2\tilde{\Delta} \sin^2\theta,
\ee
and $\tilde{a}=\tilde{J}/\tilde{M}$. Then one can find that the Einstein tensor for above metric is linear in derivatives of the mass function $\tilde{m}(r)$ and the rotational parameter $\tilde{a}$ appears in a convoluted form. This indicates that the linear decomposition of the mass function 
\be
\tilde{m}=m(r)+\alpha m_s(r),
\ee
and the assumption that the rotational parameter $\tilde{a}$ remains unaffected
will generate a linear decomposition of the Einstein tensor of the form 
\be\label{Gmunudecoupling}
\tilde{G}^\sigma_\gamma(\tilde{m},\tilde{a})=G^\sigma_\gamma(m,\tilde{a})+\alpha G^\sigma_\gamma(m_s,\tilde{a}),
\ee
where the parameter $\alpha$ is introduced to keep track of the
deformation, the mass functions $m$ and $m_s$ are generated
by  $T_{\mu\nu}$ and $S_{\mu\nu}$ in Eq.(\ref{energymomentumtenosr}). In this case,  the Einstein equation (\ref{Einsteinequation}) can split in two parts: the first part is 
 \be
 G^\mu_\nu(m,a)=0,
 \ee
whose solution is the Kerr metric with $m$ and $a$ being the corresponding parameters, and the second one is
\be
\alpha G^\mu_\nu(m_s,a)=8\pi S^\mu_\nu,
\ee
where to achieve the decoupling (\ref{Gmunudecoupling}) one has to demand $\tilde{a}=a$. The solution to above equation  has the same form as the one in Eq. (\ref{metricansatz}) with $m(r)\to \alpha m_s(r)$. Moreover, by requiring that $S_{\mu\nu}$ satisfies the  strong energy condition in the region outside the event horizon, Ref. \cite{Contreras:2021yxe} obtained the HKBH solution, whose metric in terms of the Boyer-Lindquist coordinates  can be written as
\begin{eqnarray}\label{HairKerr:Boyer-Lindquist}
ds^2&=&-{\Delta_H\over\Sigma}(dt-a\sin^2\theta\,d\phi)^2
+{\sin^2\theta\over\Sigma}\left[(r^2+a^2)d\phi-a\,dt\right]^2
\nonumber\\
&&+{\Sigma\over\Delta_H}dr^2+\Sigma d\theta^2\,,
\end{eqnarray}
here $\Sigma=$ $r^2+a^2\cos^2\theta$ and $\Delta_H=$ $r^2-2Mr+a^2+\alpha r^2 e^{-r/(M-h_0/2)}$. One can see that the distinction between above expression and Kerr metric is only reflected in the function $\Delta_H$.
The parameters $(M, a)$ represent the mass and spin of the HKBH,
and the parameters $(\alpha, h_0)$ are introduced due to the presence of surrounding additional matter such as dark matter.
$\alpha$ is the parameter to characterize the deviation from \ac{KBH} and $h_0=\alpha h$ represents a
	charge associated with primary hair. $h_0$ is bounded to $h_0\leq2M$ to ensure asymptotic flatness.
 The HKBH returns to  the standard Kerr BH when parameter $\alpha=0$ or $h_0\to 2M$. As  pointed out by  \cite{Contreras:2021yxe}, the existence of the HKBH does not contradict the conclusions of \cite{Gurlebeck:2015xpa}. This is because in  \cite{Gurlebeck:2015xpa} the validness of the no hair theorem for BH in astrophysical background requires that the external sources distribute outside the horizon. However, in \cite{Contreras:2021yxe} the fluid modifying the Kerr geometry overlaps the BH horizon. 
 
The roots of $\Delta_H=0$ are the locations of the horizons of the HKBH \eqref{HairKerr:Boyer-Lindquist},
which in general can be obtained only numerically. However, if the deviation parameter is small, i.e., $\alpha\ll1$, the roots can be found analytically, as shown in \cite{Li:2022hkq}.  In this case, the radii of the two horizons of the HKBH are given by
\bea
r_+^H &=& M+\sqrt{M^2-a^2-\alpha r_+^{I~2} e^{-r_+^{I}/(M-h_0/2)}}\\
r_-^H &=& M-\sqrt{M^2-a^2-\alpha r_-^{I~2} e^{-r_-^{I}/(M-h_0/2)}}
\eea
where  $r_{+,-}^{I}$ are the first-order solutions of the equation $\Delta_H=0$,
\bea
r_+^I &=& M+\sqrt{M^2-a^2-\alpha r_+^{I~2} e^{-r_+/(M-h_0/2)}}\\
r_-^I &=& M-\sqrt{M^2-a^2-\alpha r_-^{I~2} e^{-r_-/(M-h_0/2)}}.
\eea
Moreover, $r_\pm$ are the radii of the horizons of the Kerr BH, namely the solutions of $\Delta_H(\alpha=0)=0$. One can proceed the iteration to obtain higher-order solutions, but according to the analysis in \cite{Li:2022hkq}, the second-order solutions are sufficiently accurate.
Thus the function $\Delta_H$ is given approximately  by
\bea
\Delta_H \approx (r-r_+^H )(r-r_-^H).
\eea
\subsection{Perturbation Equations}\label{perturbeqs}
Teukolsky \cite{Teukolsky:1973ha} demonstrated  that the gravitational perturbations of the Kerr BH can be described by the
Newman-Penrose variable $\Psi_4$, which can be decomposed by two linear ordinal differential equations due to the symmetries of the
Kerr spacetime. Unfortunately, the procedure taken by Teukolsky cannot be applied directly to the HKBH, as the latter is not a solution of the vacuum Einstein equations. However, Ref.\cite{Li:2022hkq} noticed the similarity between the HKBH and the KN BH, that is, the distinction of the metric is only embodied through the function $\Delta_H$. It is known that if the electromagnetic perturbations are ``frozen'', the gravitational perturbations of the KN BH can be captured well by the Dudley-Finley equation \cite{Dudley:1978vd,Kokkotas:1993ef,Berti:2005eb}.\footnote{However, if the electromagnetic perturbations need to be taken into account, even the electric charge of the KH BH is very small, the Dudley-Finley equation becomes inaccurate \cite{Mark:2014aja}.} Therefore, one expects that if the perturbations of the matter content are ignored, the  gravitational perturbations of the HKBH can be described by the Dudley-Finley equation with the replacement $\Delta_{KN}\to\Delta_{H}$. Since the specific property of the matter content is blind to us, we take the strategy to ignore its perturbations.\footnote{The perturbations from the surrounding fluid of a spherically symmetric hairy BH were studied in \cite{Cardoso:2022whc}. } In this case, the  radial part of the perturbation equations can be written as
\begin{equation}
\Delta_H ^{2}\frac{d}{dr}\left(\frac{1}{\Delta_H}\frac{d R_{\ell m \omega}}{dr}\right) - V(r)R_{\ell m \omega} =
-\mathcal{T}_{\ell m \omega} \label{Teuko:eq},
\end{equation}
and
\begin{equation}
V(r) = -\frac{K^2 + 4 i(r-M)K}{\Delta_H} + 8i\omega r + \lambda + a^2\omega^2 -2a m\omega-2,
\end{equation}
where $V(r)$ is the effective potential function, $K=(r^2+a^2)\omega - a m $,
$\lambda$ is the eigenvalue of spheroidal harmonic $_{-2}S^{a\omega}_{\ell m}$ and $\mathcal{T}_{\ell m \omega}$ is the source term.
The angular equation is of the following form
\begin{eqnarray}\label{Teuko:eq:angular}
& &\Bigl[{1 \over \sin\theta} \frac{d}{d\theta}\left(\sin\theta {d \over d\theta}\right)
-a^2\omega^2\sin^2\theta
 -{(m-2\cos\theta)^2 \over \sin^2\theta}
\nonumber\\
& &~~~~~~~~~~~~~~~
+4a\omega\cos\theta-2+2ma\omega+\lambda\Bigr]
{}_{-2}S_{\ell m}=0.
\end{eqnarray}
which keeps same with the angular part of perturbation equations  in the Kerr spacetime \cite{Teukolsky:1973ha}.

There are two sets of independent homogeneous solutions to the radial equation \eqref{Teuko:eq},
which are written as the following
\begin{equation} \label{asymptoticsin}
R^{\rm in}_{\ell m \omega} \sim
\begin{cases}
 \displaystyle
B^{\rm trans}_{\ell m \omega} \Delta^2 e^{-i k r_*} & \text{ as } r_* \to - \infty\\
 \displaystyle
 r^3 B^{\rm out}_{\ell m \omega}  e^{i \omega r_*}  +  r^{-1} B^{\rm in}_{\ell m \omega} e^{- i \omega r_*} & \text{ as } r_* \to
+ \infty
\end{cases} \,,\\
\end{equation}
\begin{equation} \label{asymptoticsup}
R^{\rm out}_{\ell m \omega} \sim
\begin{cases}
 \displaystyle
 C^{\rm out}_{\ell m \omega} e^{i k r_*}  +  \Delta^2 C^{\rm in}_{\ell m \omega} e^{-i k r_*} & \text{ as } r_* \to - \infty \\
 \displaystyle
 r^3 C^{\rm trans}_{\ell m \omega} e^{i \omega r_*} & \text{ as } r_* \to + \infty \\
\end{cases} \,,
\end{equation}
where $k =\omega-am/(2M r_+^H) $, and $r_\ast$ is the tortoise coordinate defined by $dr_\ast/dr=(r^2+a^2)/\Delta_H$.

Since there is the long-ranged potential in Eq. \eqref{Teuko:eq},
the solutions \eqref{asymptoticsin}-\eqref{asymptoticsup} are convergent poorly at infinity. To overcome this problem,
Sasaki and Nakamura developed a suitable scheme that the Teukolsky equation \eqref{Teuko:eq} is transformed into
SN equation with a  short-ranged potential \cite{Sasaki:1981sx}.
Following the method in Ref. \cite{Gralla:2015rpa}, we compute the boundaries conditions and solve the SN equation,
then get the homogeneous solutions of Eq. \eqref{Teuko:eq} and energy fluxes at infinity and near the horizon.
In order to obtain spin-weighted spheroidal harmonics and eigenvalue $\lambda$,
we make use of Black ~ Hole ~Perturbation ~Toolkit ~package \cite{BHPToolkit}.

\subsection{Adiabatic Evolution}\label{evolution}
In this subsection we introduce the procedure of evolving inspiral orbital parameter under the adiabatic approximation condition.
Concretely, the motion of \ac{CO} is kept in the circular orbits on the equatorial plane near the HKBH \eqref{HairKerr:Boyer-Lindquist},
and its circular orbital period is far less than the inspiral time-scale due to the huge different mass ratio, we can make the hypothesis that the energy loss of EMRI system results from the radiant GW.

Following Refs. \cite{Saijo:1998mn,Jefremov:2015gza,Piovano:2020zin}, for the circular orbits on the equatorial plane the orbital energy $E$ and the angular momentum $L_z$ of \ac{CO} can be given by
\begin{align}
\hat{E} &=\frac{E}{\mu} =\frac{\hat{r} \sqrt{\Delta_H} +\hat{a} U_\mp }{\hat{r} \sqrt{1 - U_\mp^2}} \
,\label{eq:energyofr}\\
\hat{L}_z &=\frac{L_z}{\mu M} =\frac{ \sqrt{\Delta_H} \hat{a}  +  (\hat{r}^3 + \hat{r}\hat{a}^2 ) U_\mp}{\hat{r}^2 \sqrt{1 - U_\mp^2}} \ .\label{eq:jzofr}
\end{align}
where we introduced the hatted dimensionless quantities as $\hat{a}=a/M$ and $\hat{r}=r/M$,
and the expression $U_\mp$ is given by \cite{Piovano:2020zin}
\begin{equation}\label{eq:Ump}
U _\mp = -\frac{ \hat{a}\hat{r}^3 \mp \sqrt{\hat{r}^7}}{\sqrt{\Delta_H}\hat{r}^3 } \ .
\end{equation}
Here the minus and plus sign in Eq.~\eqref{eq:energyofr}-\eqref{eq:Ump} denote prograde and retrograde orbits, respectively. The orbital angular frequency for the orbits is written as \cite{Piovano:2020zin}
\begin{equation}
\frac{d\phi(t)}{dt} = \Omega(t) =\frac{1}{M} \frac{ \hat{a}  +\sqrt{\Delta_H}U _\mp}
{ (\hat{r}^2+\hat{a}^2)\hat{a}^2  +\hat{a}\sqrt{\Delta_H}U _\mp } \ . \label{eq:omegaofr}
\end{equation}
In the frame of adiabatic approximation, we can evolve the orbital radius $\hat{r}_0(t)$ by the flux balance equation
\begin{equation}
\frac{d \hat{r}_0(t)}{dt} =- \dot{E}_{\rm GW}\left(\frac{d\hat{E}}{d\hat{r}}\right)^{-1}\Bigg{|}_{\hat{r}=r_0} \;, \label{eq:balance}
\end{equation}
where $\dot{E}_{\rm GW}$ is the total energy flux produced by EMRI sysytem,
which consists of the energy fluxes at the infinity $\dot{E}^{\rm \infty}_{\rm GW}$ and near the horizon $\dot{E}^{\rm H}_{\rm GW}$,
their expressions are of the following form~\cite{Hughes:2001jr}
\bea
\dot{E}^{\rm \infty}_{\rm GW} = \sum_{\ell,m}
{\Bigl|Z^\infty_{\ell m\omega}\Bigr|^2\over2\pi (m\Omega)^2} \ , \label{eq:energyfluxinf} \\
\dot{E}^{\rm H}_{\rm GW}  =\sum_{\ell,m}
{ \alpha_{\ell m} \Bigl|Z^{\rm H}_{\ell m\omega}\Bigr|^2\over2 \pi (m\Omega)^2} \ , \label{eq:energyfluxhor}
\eea
where $\alpha_{\ell m}$ is given by Ref.~\cite{Hughes:2001jr}, and
\begin{align}
Z^\infty_{\ell m \hat{\omega}}  &=  \frac{1}{2i \omega B^{\textup{in}}_{\ell m\omega}}   \int_{\hat{r}_+}^{\infty} \dd
\hat{r}'\frac{\Rout(\hat{r}')
}{\Delta_H^2}\mathcal{T}_{\ell m \omega}(r') \ , \label{eq:infamp}\\
Z^H_{\ell m \omega} &= \frac{B^{\textup{tran}}_{\ell m \omega}}{2i \omega B^{\textup{in}}_{\ell
m\omega}C^{\textup{tran}}_{\ell m\omega}}
\int_{\hat{r}_+}^{\hat{r}}\dd
\hat{r}'\frac{\Rin(\hat{r}')}{\Delta_H^2} \mathcal{T}_{\ell m\omega}(\hat{r}') \ , \label{eq:horamp}
\end{align}
where the full expression of $\mathcal{T}_{\ell m \omega}(r')$ is given by ~\cite{Hughes:2001jr}.

For dominant mode of GW, the phase is related to the orbital phase $\Phi_{\rm GW}=2\phi$.
To assess the difference of GW phase form EMRI with hairy and Kerr BH,
we compute the dephasing  by the following formula:
\bea
\delta \Phi(t) = \Phi^{\rm Kerr}_{\rm GW}(t) -  \Phi^{\rm Hair}_{\rm GW}(t).
\eea
Here $\Phi^{\rm Hair}_{\rm GW}(t)$ and $\Phi^{\rm Kerr}_{\rm GW}(t)$ are the GW phase in the spacetime of  hairy and Kerr BH,
respectively.

In this paper we solve Teukolsky equation for each $(\ell,m)$ mode up to $\ell_{\rm max}=10$, and calculate the energy fluxes by summing all $(\ell,m)$ mode with Eq. \eqref{eq:energyfluxinf}-\eqref{eq:energyfluxhor}.
With the energy fluxes at hand,  we can solve the evolution equation \eqref{eq:balance} using the Euler method, and calculate the trajectory of orbital radius for the circular and equatorial orbits.
It should be noted that the trajectory of the secondary body is evolved adiabatically from $r_0=10M$ to $r_{\rm ISCO}+\delta M$,
where $\delta M = 0.01M$  and $r_{\rm ISCO}$ is \ac{ISCO} in the spacetime of \ac{HKBH} and its location is investigated in Ref. \cite{Wu:2023wld}.

\subsection{Waveform Analysis}\label{waveformanalysis}
In this subsection we introduce the method of waveform calculation and analysis.
Firstly, we obtain trajectory of orbital radius in term of the aforementioned way, then substitute into
the waveform formula, the waveform emitted from the EMRI at infinity is computed \cite{Hughes:2001jr} with these function
\begin{eqnarray}
 h_+- i h_\times =& \displaystyle-\frac{2}{\sqrt{2 \pi}} \sum_{\ell,m}
\frac{Z_{\ell m \omega}^{\infty}}{(m\Omega)^2}
e^{i  \omega\left( r_*- t \right)} \nonumber \\
&\times ~_{-2}S_{\ell m \omega}(\vartheta) e^{i m \varphi} \,, \label{waveform}
\end{eqnarray}
where $\vartheta$ is the angle between the  line of sight of the observer and the spin axis of the MBH,
and $\varphi\equiv\phi(t=0)$.

In order to evaluate the difference of GW waveforms from \ac{KBH} and \ac{HKBH} with space-borne GW detectors quantitatively,
we compute a function  mismatch $\mathcal{M}$ of two kinds of waveforms.
For two waveforms $h_a(t)$ and $h_b(t)$, mismatch $\mathcal{M}$ is related to overlap by $\mathcal{O}=1-\mathcal{M}$,
and the overlap $\mathcal{O}$ is given by
\be
\mathcal{O}(h_a|h_b) = \frac{<h_a|h_b>}{\sqrt{<h_a|h_a><h_b|h_b>}}
\ee
with the noise-weighted inner product $<*|*>$ is defined by
\begin{align}\label{inner}
<h_a |h_b > =2\int^\infty_0 df \frac{\tilde{h}_a^*(f)\tilde{h}_b(f)+\tilde{h}_a(f)\tilde{h}_b^*(f)}{S_n(f)},
\end{align}
where the quantities with tilde represent the Fourier transform of waveform data,
the star sign means complex conjugation,  and $S_n(f)$ is noise power spectral density of LISA~\cite{LISA:2017pwj}.
A rule of thumb is proposed,
in which the GW detector can distinguish two kinds of waveforms when the mismatch $\mathcal{M}\geq D/(2\rho^2)$,
here $D$ and $\rho$ is the number of intrinsic parameters and \ac{SNR} of the GW signal \cite{Flanagan:1997kp,Lindblom:2008cm}.
In this paper the minimum value of mismatch that is recognized by LISA $\mathcal{M}_{\rm min}=0.0075$ if
\ac{SNR} of the GW signal is assumed as $\rho=20$.
\section{Result}
In this section we present some results about waveform from \ac{HKBH}, trajectory and mismatch.

\begin{figure*}[th]
 \centering
 \includegraphics[width=0.49\textwidth]{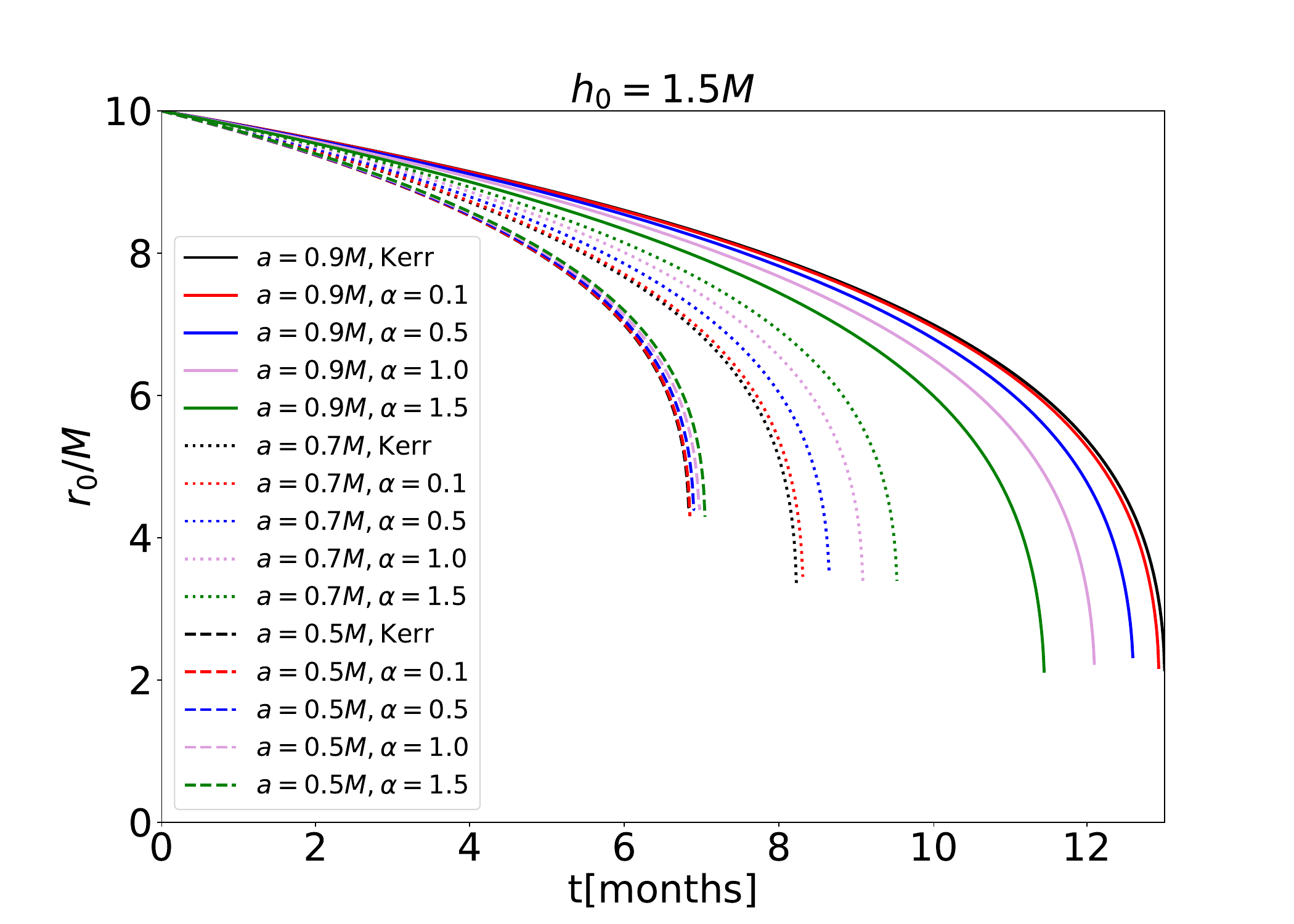}
 \includegraphics[width=0.49\textwidth]{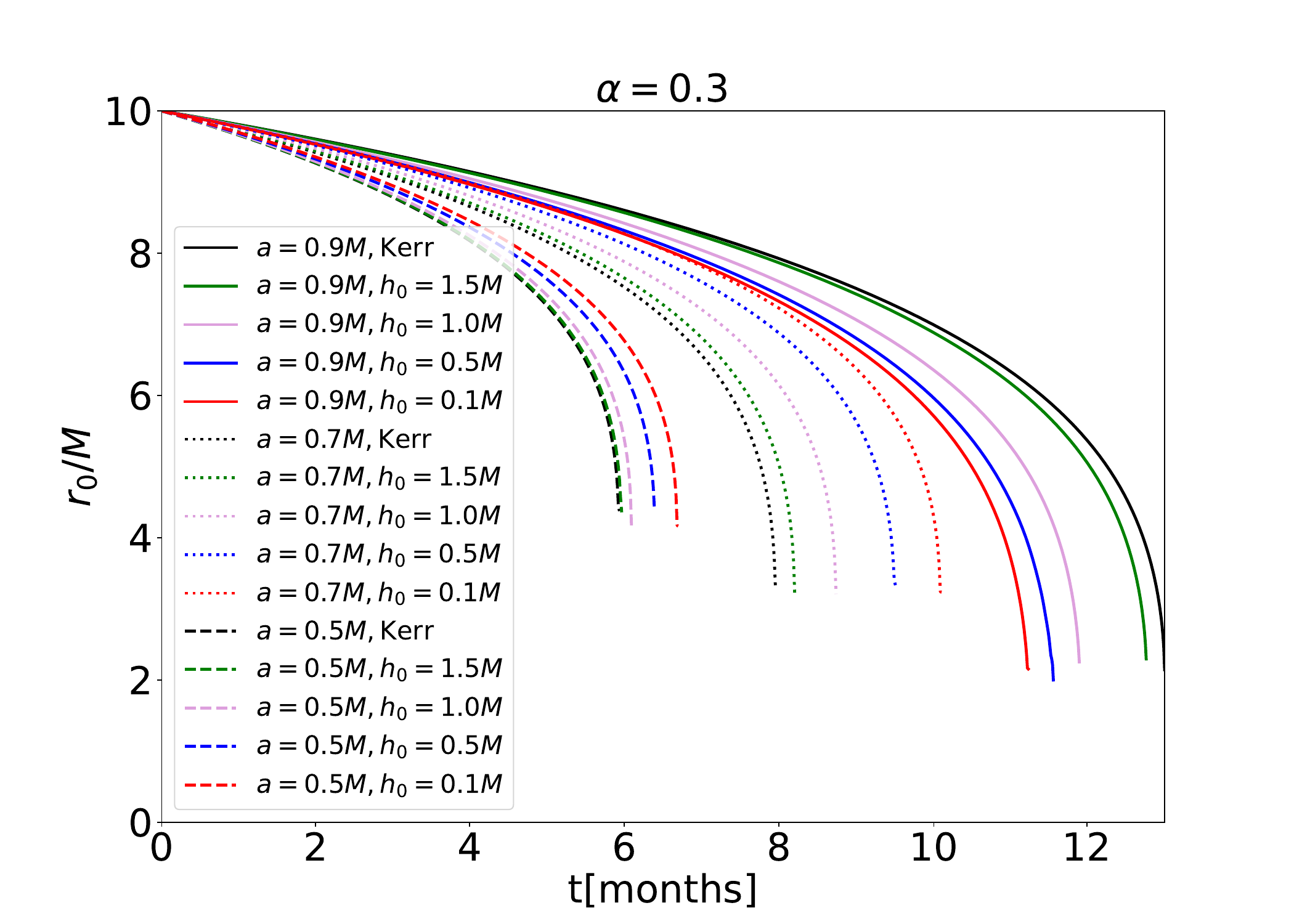}
 \caption{Comparison of trajectories for EMRI with Kerr BH and hairy Kerr BH
 for fixed $h_0=1.5M$ (left panel) and $\alpha=0.3$ (right panel) is plotted, which includes different spins, hair and deviation parameters. The evolution of circular orbits ranges in the radii from  $r_0=10M$ to \ac{ISCO}.
  }\label{traj:Nonkerr}
\end{figure*}

 We choose the EMRI system with $M=10^6\Msun$ for the central BH and $\mu=10\Msun$ for the small CO. The evolution of the CO's trajectory under the radiation reaction  with various  spins and values of the parameters of the HKBH is shown in  Fig. \ref{traj:Nonkerr}. The left panel describes the influence from the values of the deviation parameters for a given hair charge $h_0=1.5M$ while in the right panel we let the deviation parameter be fixed as $\alpha=0.3$ and the hair charge vary.
From left panel of Fig. \ref{traj:Nonkerr},
it is obvious that the distinction between the trajectory of the HKBH from that of the KBH grows significantly
as $\alpha$ increases. Moreover, the spin of the central BH enhances the distinction.
On the other hand, from the right panel, the difference between the trajectories in the spacetime of \ac{HKBH} and \ac{KBH} becomes more and more prominent as the hairy charge $h_0$ becomes smaller.
The reason is that the function $\Delta_H$ is a monotonic decreasing function of $h_0$.
As a result, a larger  $h_0$ leads to a smaller deviation from the KBH.
As we will see in the following, the similar phenomena occur  for the dephasing and mismatch of the EMRI waveforms.

\begin{figure*}[th]
 \centering
 \includegraphics[width=0.45\textwidth]{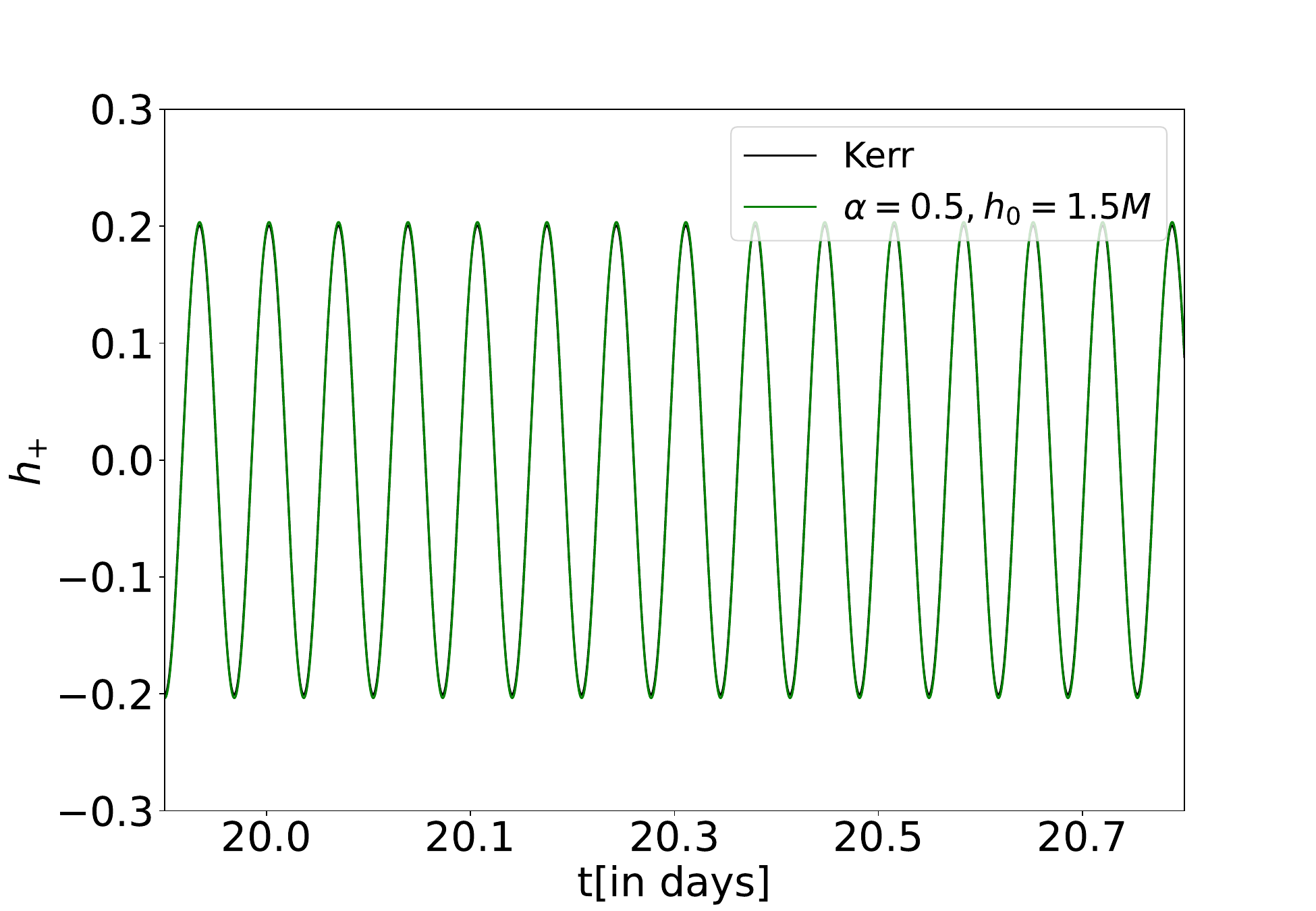}
 \includegraphics[width=0.45\textwidth]{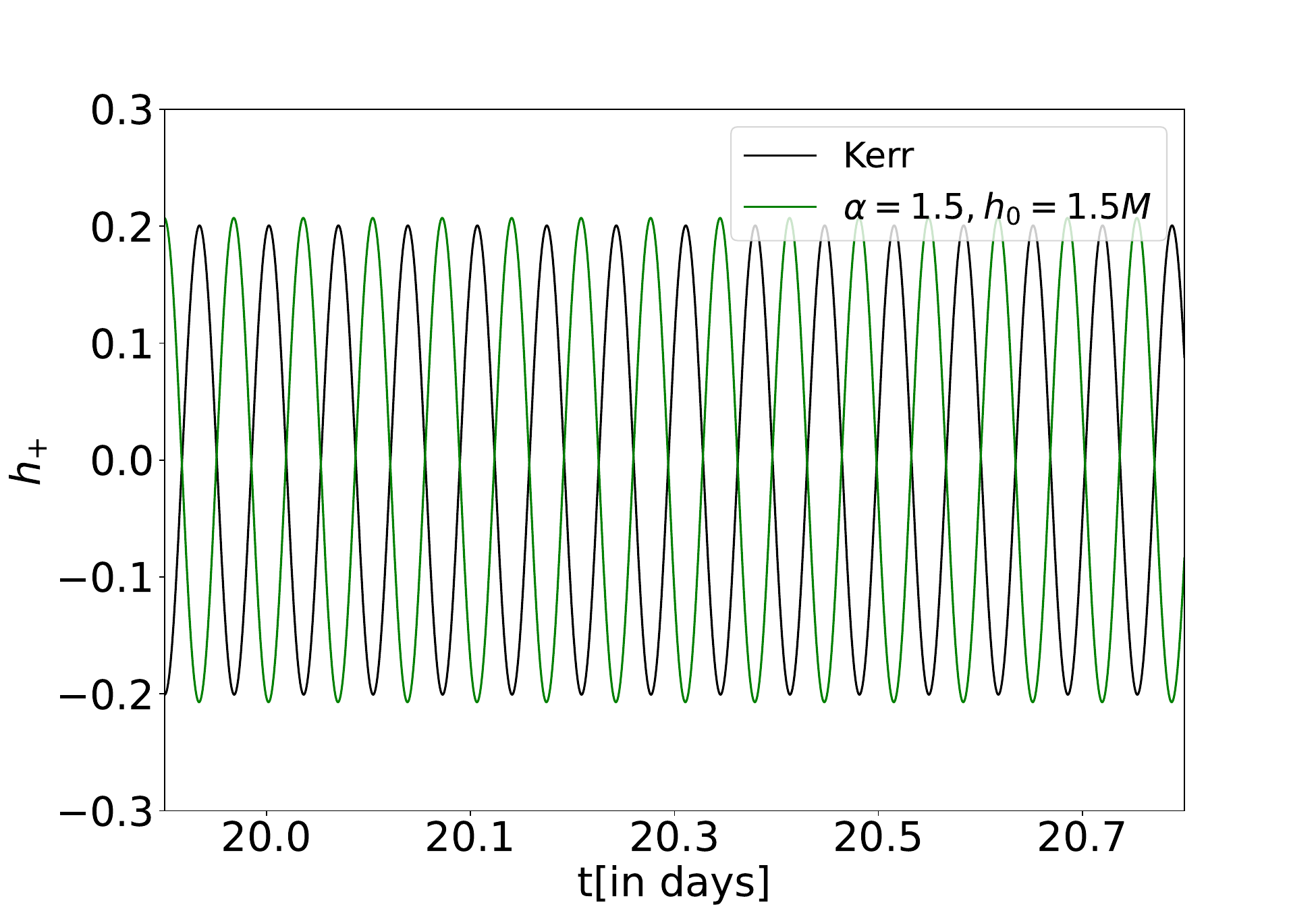}
 \caption{Comparison of waveforms from EMRI with the \ac{KBH} and \ac{HKBH}  for $r_0=10M$ and spin $a=0.9M$,
 where the parameters of \ac{HKBH} are $\alpha=0.5, h_0=1.5M$ (left  panel) and $\alpha=1.5, h_0=1.5M$ (right panel).
 For the bigger  deviation parameter $\alpha$, EMRI with \ac{HKBH} results in more dephasing comparing with the \ac{KBH} case.
    }\label{wave:Kerr:NonKerr}
\end{figure*}

In Fig. \ref{wave:Kerr:NonKerr} the EMRI waveforms as functions of time with $h_0=1.5M$ and spin $a=0.9M$ are plotted,
where $\alpha=0.5$  and $\alpha=1.5$ in the left and right panels, respectively.
The inspiral with  initial phase $\phi(0)=0$ and initial orbital radius $r_0=10M$ lasts for about 20 days.
As shown in this case, the dephasing of waveform becomes apparent with a larger deviation parameter $\alpha=1.5$. This is in accordance with its effect on the orbital evolution. Therefore, we expect that for a given $\alpha$ a smaller $h_0$ would yield  more prominent distinction between the two waveforms of the HKBH and KBH.


\begin{figure*}[th]
 \centering
  \includegraphics[width=0.49\textwidth]{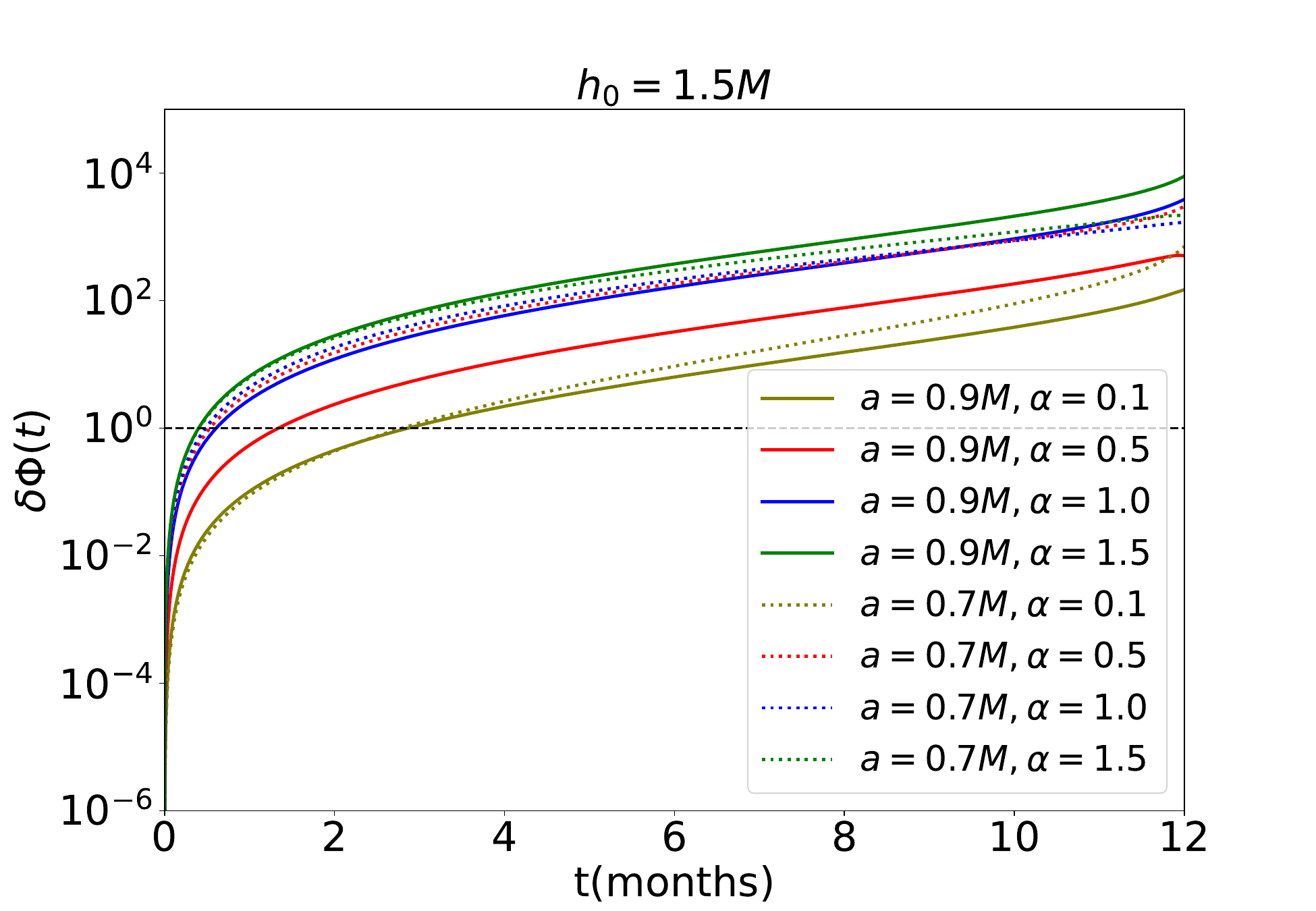}
  \includegraphics[width=0.49\textwidth]{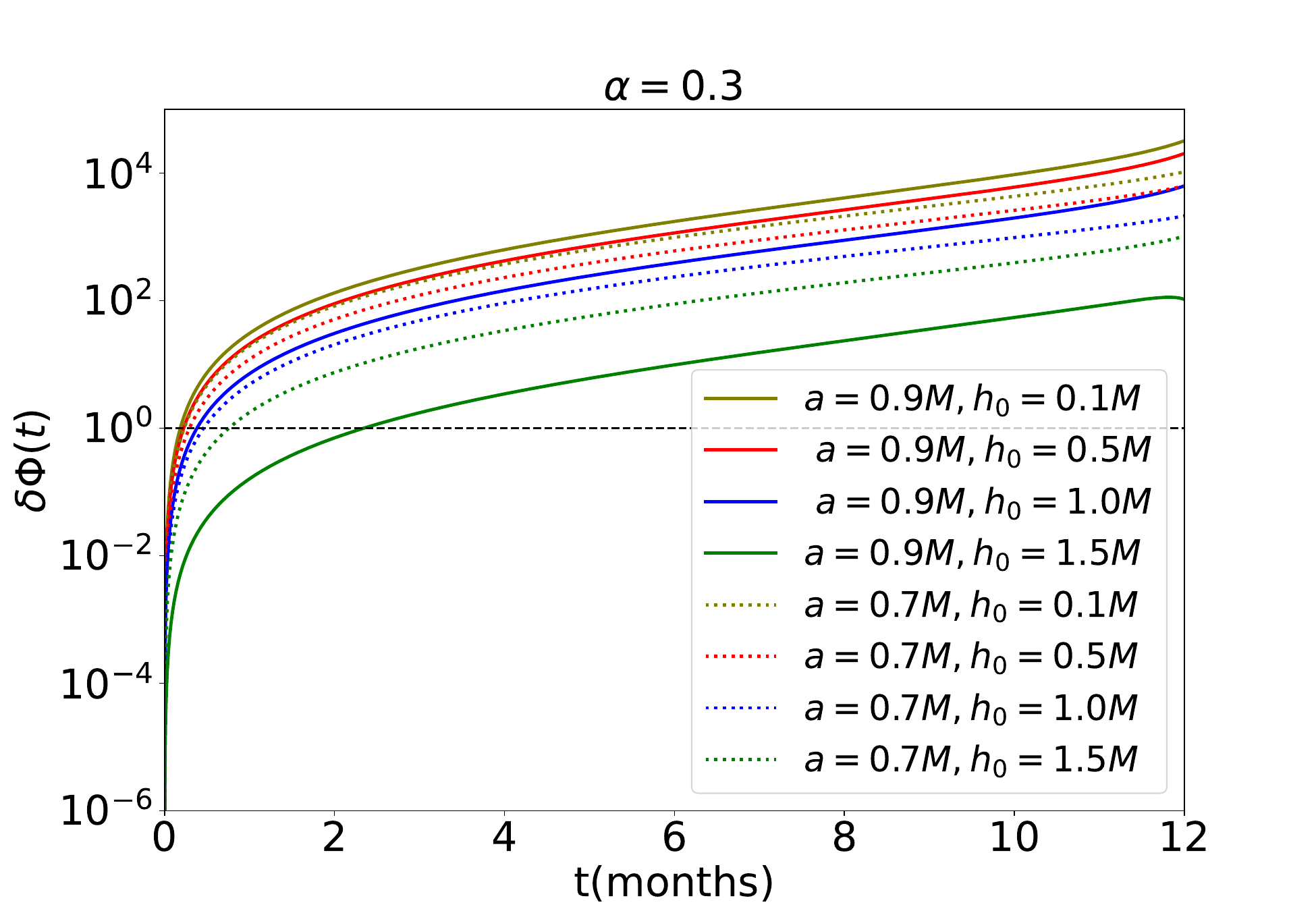}
 \caption{Dephasing of waveform form EMRI with Kerr BH and hairy rotating BH
  for $h_0=1.5M$ (left panel)  and $\alpha=0.3$ (right panel) is depicted,  which includes the parameters $h_0 \in(0.1,0.5,1.0,1.5),a\in(0.7M, 0.9M)$ in the left panel and  $\alpha\in(0.1M,0.5M,1.0M,1.5M),a\in(0.7M, 0.9M)$ in the right panel, respectively.
  }\label{deltaphi:NonKerr}
\end{figure*}
To study the effects from various parameters on the detection capability of space-based GW detectors for the EMRIs of the HKBH, we show  the difference of waveform phase as a function of observation time  in Fig. \ref{deltaphi:NonKerr}
The parameters associated with the HKBH are set as $\alpha\in(0.1,0.5,1.0,1.5), h_0=1.5M$ in the left panel and $h_0\in(0.1M,0.5M,1.0M,1.5M),\alpha=0.3$ in the right panel, respectively.
The black and dashes horizontal line in Fig. \ref{deltaphi:NonKerr} is the threshold $\delta \Phi\sim 1\rm rad$, above which the two signals can be distinguished by space-based GW detectors. This provides a rather rough
constraint on the parameters related with the HKBH, and the validity of such criterion is argued in Ref. \cite{Datta:2019epe}.
When the effect of the additional  parameters of the \ac{HKBH} results in a dephasing $\delta \Phi \gtrsim 1\rm rad$,
the HKBH has the potential to be detected by LISA \cite{Lindblom:2008cm}.
From Fig. \ref{deltaphi:NonKerr}, we can see that for observation longer than three months, the accumulated dephasing can easily exceed the threshold of $1$ rad, with the parameters $\alpha$ is even as small as $0.1$ and $h_0$ is as large as $1.5$. Moreover, the dephasing becomes evident with the increase of $\alpha$  and the decrease of $h_0$. Thus, the effect from the hair of the HKBH on the dephasing is the same as it has on the orbital evolution.
\begin{figure*}[th]
 \centering
 \includegraphics[width=0.49\textwidth]{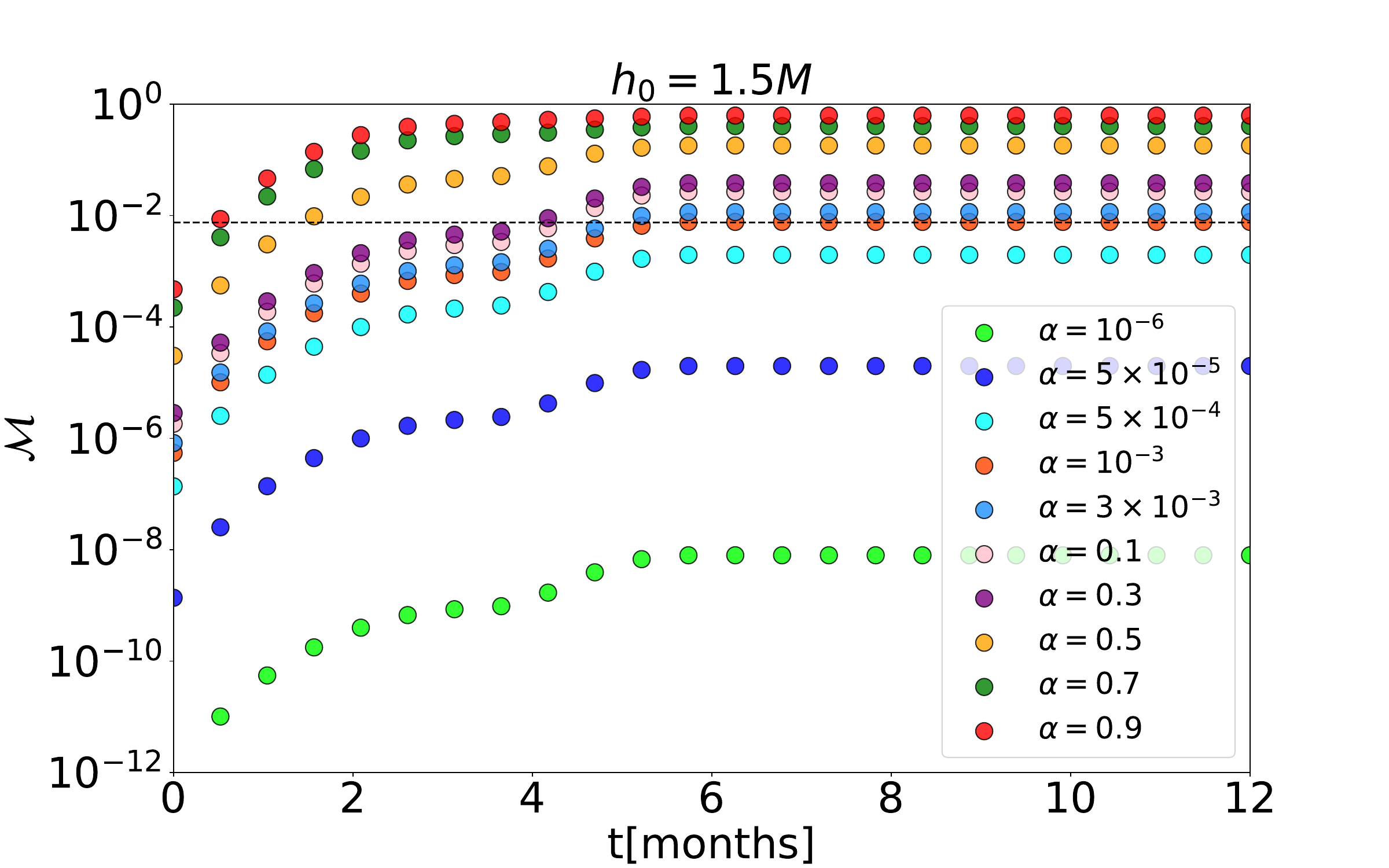}
  \includegraphics[width=0.49\textwidth]{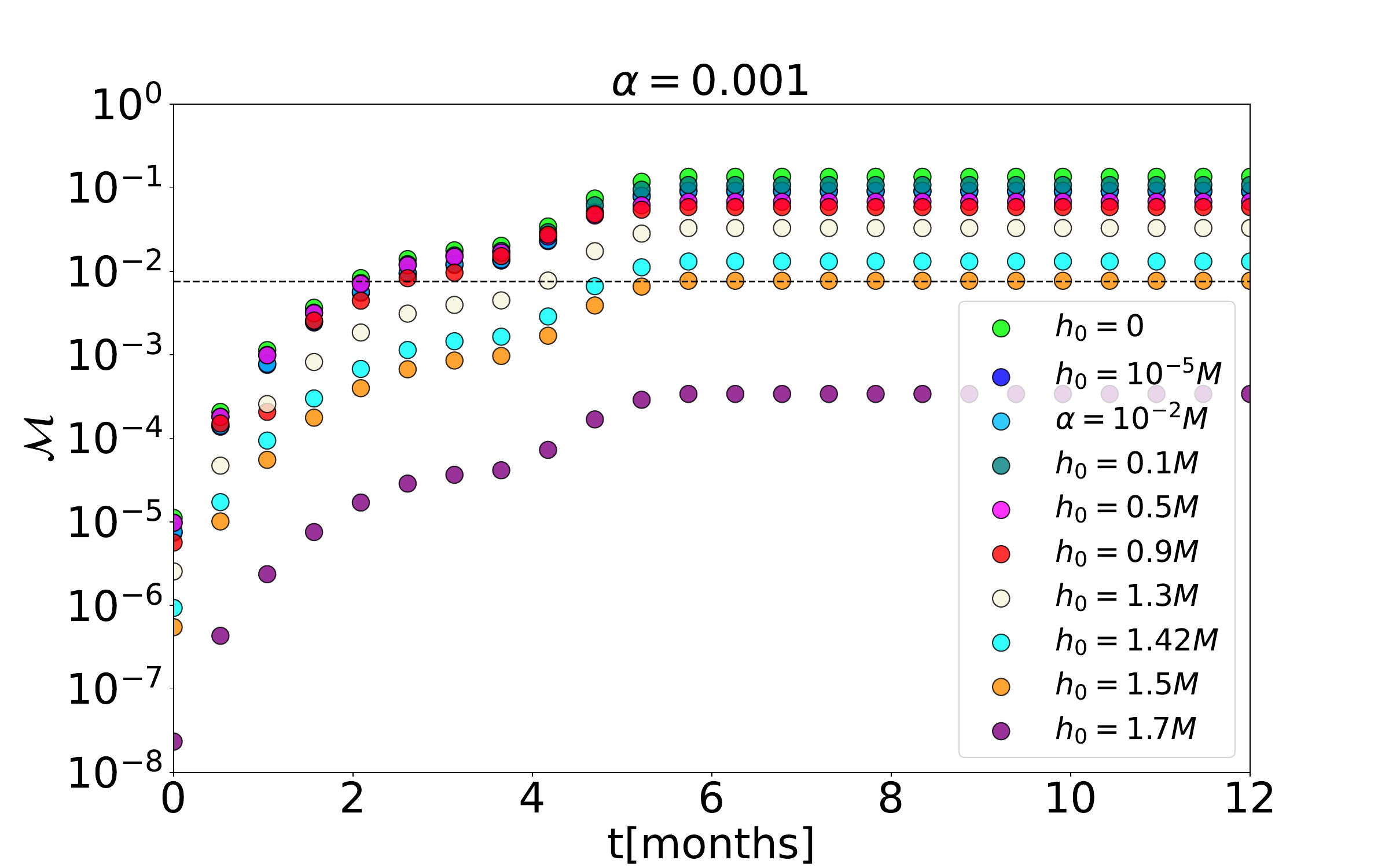}
 \caption{Mismatch as a function of observation time for EMRI waveforms from \ac{KBH} and \ac{HKBH} is depicted,
 where the parameter are set as $h_0=1.5M,\alpha\in(10^{-6},5\times10^{-5},5\times10^{-4},10^{-3},3\times10^{-3},0.1,0.3,0
 5,0.7,0.9)$ in the left panel and $\alpha=0.001,h_0\in(0, 10^{-5}M,10^{-2}M,0.1M,0.3M,0.5M,0.9M,1.3M,1.42M,1.5M,1.7M)$ in the right panel, respectively.
  }\label{Mismatch:Contourf}
\end{figure*}

To quantitatively evaluate the ability of space-based gravitational wave detectors like LISA to discern the EMRI waveforms of the HKBH, we calculate the mismatch between the waveforms of the HKBH and KBH, taking into account various parameters and as a function of observation time. The parameters of the central BH are fixed as $M=10^6\Msun, \mu=10\Msun$ and $a=0.9M$. The results with various values of the deviation parameters $\alpha$ and hair charge $h_0$ are plotted in Fig. \ref{Mismatch:Contourf}.

The black and dashes horizontal line represents for the threshold value $\mathcal{M}_{\rm min}=0.0075$  of mismatch, above which the HKBH can be distinguished by LISA.
From the left panel of Fig. \ref{Mismatch:Contourf}, the mismatch becomes larger gradually as the deviation parameter $\alpha$ increases for a fixed $h_0=1.5M$. In this case, we find that the threshold value of deviation parameter is $\alpha_{\rm min} \approx 0.001$ for 1 year observation by LISA.
From the right panel of Fig. \ref{Mismatch:Contourf}, the mismatch is less sensitive to the hair charge $h_0$ and the threshold value is $h_0^{\rm  min} \approx 1.5$ for the fixed parameter $\alpha=0.001$.
Specifically, the mismatch varies several times in range of $h_0\in[0,1.5M]$.

\section{Conclusion}\label{Conclusion}
EMRI, as a peculiar binaries system, has attracted extensive attention with the studies of detectability sources and testing gravitational theories using the future space-based detectors.
A rotating black hole with primary hair called hair Kerr BH (HKBH) was constructed  resorting to the gravitational decoupling approach \cite{Contreras:2021yxe}.
The parameters of the HKBH are expected to be measured accurately with the observation of the GWs from the EMRI \cite{Babak:2017tow,Fan:2020zhy,Zi:2021pdp}.
In this paper we calculated the EMRI waveforms emitted from the \ac{HKBH} with the modified Teukolsky equation, and the requirement that the  waveforms of the \ac{HKBH} and \ac{KBH} can be distinguished by LISA  placed a rough constrain of the additional parameters.

Firstly, we evolved the circular orbital radius under radiation reaction and computed the time domain waveforms from EMRI systems of \ac{HKBH} and \ac{KBH}, respectively. The trajectories of the \ac{CO} with various parameters were shown in Fig. \ref{traj:Nonkerr} and the
 two kinds of EMRI waveforms from \ac{KBH} and \ac{HKBH} in the time domain were displayed  in Fig. \ref{wave:Kerr:NonKerr}. Then, we calculated the dephasing and the mismatch of the  two kinds of EMRI waveforms to assess the detectability of the HKBH by LISA, as shown in Fig. \ref{deltaphi:NonKerr} and Fig. \ref{Mismatch:Contourf}. It was found that for 1 year observation, the HKBH  with the deviation parameter $\alpha_{\rm min}\sim 0.001$ and  hair charge $h_0^{\rm min}\sim 1.42M$ can be discerned by LISA.

In this paper we only considered the equatorial, circular orbits, however, it is expected that our results would be of the same order in magnitude with those of the eccentric and inclined trajectories.
Nevertheless, it would be interesting to study the GWs emitted from the generic orbits of EMRI around \ac{HKBH} to obtain reliable conclusion. Additionally, the perturbation equation in this paper was taken as
Dudley-Finley equation \cite{Kokkotas:1993ef,Berti:2005eb} under the condition that the perturbations from the matter content are ignored. This results in degeneracy of the two parameters $\alpha$ and $h_0$, since they as a whole appear only in a single form in $\Delta_H$. If the matter content of the HKBH is explicitly given, then its perturbations can be used to break the degeneracy. This would make it possible to uniquely constrain each parameter with the observation of EMRI signals. A significant difference between Kerr BH and alternatives is that the tidal Love number of the former is exactly zero \cite{Charalambous:2021kcz,Chia:2020yla,LeTiec:2020spy} but that of the latter is usually not. Therefore, with GW observations, the measurement of the tidal Love number could be used to judge whether the object is Kerr BH or the HKBH.

\begin{acknowledgments}
	This work makes use of
	the Black Hole Perturbation Toolkit package. The work is in part supported by NSFC Grant
	No.12205104 and the startup funding of South China University of
	Technology.
\end{acknowledgments}

%

\end{document}